\documentclass[12pt,preprint]{aastex}
\usepackage{graphicx}
\usepackage{epsfig}
\begin{document}

\title{Neutrino Emission and Mass Ejection in Quark Novae}

\author{Petteri Ker\"anen$^1$}
\affil{$^1$Nordic Institute for Theoretical Physics, Blegdamsvej 17,
DK-2100 Copenhagen \O , Denmark}
\email{keranen@nordita.dk}
\author{Rachid Ouyed$^2$}
\affil{$^2$Department of Physics and Astronomy, University of Calgary,
Calgary, Alberta, Canada}
\email{ouyed@phas.ucalgary.ca}
\and 
\author{Prashanth Jaikumar$^3$}
\affil{$^3$Physics Department, McGill University, 3600 University Street, Montreal H3A 2T8, Qu\'ebec, Canada}
\email{jaikumar@hep.physics.mcgill.ca}


\begin{abstract} We explore the role of neutrinos in a Quark Nova explosion.
We study production of neutrinos during this event, their propagation
and their interactions with the surrounding quark matter and the
neutron-rich envelope. We address relevant physical issues such as
the timescale for the initial core collapse, the total energy emitted
in neutrinos and their effect on the low density matter surrounding
the core. We find that it is feasible that the neutrino burst can lead
to significant mass ejection of the nuclear envelope.  
\end{abstract}


\section{Introduction}

At high baryon density and vanishing pressure, the ground state of
bulk matter may not be the most stable isotope of iron (Fe$^{56}$),
but deconfined strange quark matter (SQM) made up of up, down and
strange $(u,d,s)$ quarks (Itoh 1970; Bodmer 1971; Witten 1984;
Farhi \& Jaffe 1984). 
In that case, once
the density for a transition to the $(u,d,s)$ phase is reached in the
core of a neutron star (NS), the entire star is contaminated and
converted into a {\sl (u,d,s)} star (Haensel et al. 1986;
Alcock et al. 1986; Olinto 1987; Olesen \& Madsen 1991; Heiselberg, Baym, \& Pethick 1991; Glendenning 1992; Cheng \& Dai 1996; Anand et al. 1997; Dey et al. 1998; 
Bombaci \& Datta 2000).  The conversion could also happen via
clustering of $\Lambda$-baryons, causing formation of clumps of
strange matter, or by seeding and neutrino sparking (Alcock, Farhi \& 
Olinto (1986)), although details of such processes are still
debated.

\begin{figure}[t!]
\centerline{\includegraphics[width=0.5\textwidth]{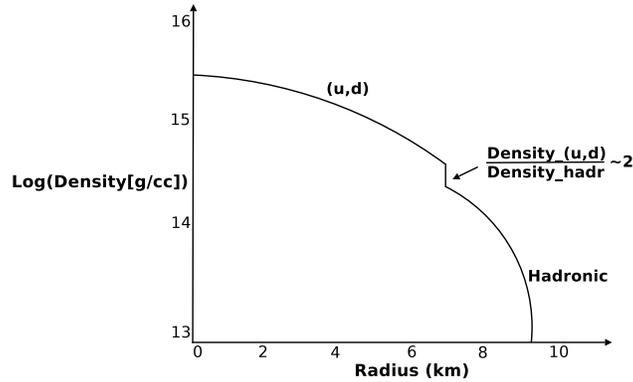}}
\caption{{\bf Schematic illustration of Density vs Radius in a hybrid star}:
A discontinuity in density, corresponding to the difference between the
hadronic and the {\it (u,d)} core occurs at the radial coordinate where the
pressure is equal to that of the mixed phase.  (Glendenning 1992;
Heiselberg et al. 1993. Also, see Chapter 9 in Glendenning (1997) for more details.}
\label{fig1}
\end{figure}

Here we focus on models where the NS core is
already converted into $(u,d)$-quark matter 
 (Alcock et al 1986). Deconfinement might occur
during or after supernova explosion
provided the central density of the proto-neutron star
is high enough to induce phase conversion
(e.g., Dai, Peng, \& Lu 1995; Xu, Zhang, \& Qiao 2000). 
Figure~\ref{fig1} is a schematic representation of the
resulting hybrid star (HS).
It is essential that there be a first-order phase transition
 between neutron matter and $(u,d)$-quark matter at 
some critical pressure when the latter becomes favored. Without
the phase transition, there will be no conversion. 

\begin{figure}[b!]
\centerline{\includegraphics[width=0.5\textwidth]{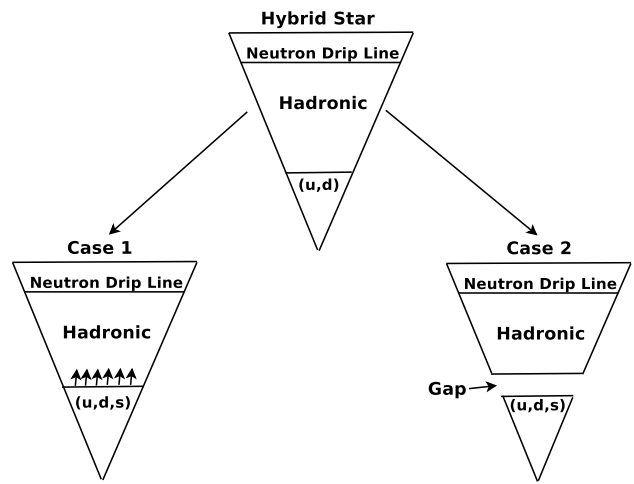}}
\caption{{\bf Conversion of a hybrid star into a {\it (u,d,s)} star}: Illustrated
are the two possible cases.
In the scenario to the right (the QN picture), the {\it (u,d)} core will
shrink faster than the hadronic envelope has time to adjust.
In the scenario to the left, the entire HS
is converted into {\it (u,d,s)} without separation of the core
and the hadronic envelope.
}
\label{fig2}
\end{figure}

In the hypothetical scenario which we call Quark-Nova (QN) (Ouyed, Dey
\& Dey 2002; ODD) the {\it (u,d)} core of the HS shrinks to the
corresponding stable, more compact {\it (u,d,s)} configuration faster
than the overlaying material (the neutron-rich hadronic envelope;
Figure~\ref{fig2}) can respond, leading to a ``gap-like'' region of
much lower density in between. In this paper, we investigate if it is
possible for the neutrinos created in the conversion\footnote{Neutrino
bursts from these conversions
have already been investigated in the literature in the
context of cold neutron stars and in supernova cores (e.g.,
Anand et al. 1997 and references therein). In the QN scenario 
 the shrinking {\it (u,d,s)} core is small enough for neutrinos to escape.
As such, our work is an extension
of previous studies to include the effects of escaping
neutrinos on the overlaying/infalling neutron-rich matter.} of neutron matter
to SQM to power the ejection of part of the neutron-rich overlaying
envelope and the suspended nuclear matter crust of the quark core.
Among the issues to consider: (i) Are the neutrinos trapped, and if
so, for how long? Since it is the conversion from {\it (u,d)} to {\it
(u,d,s)} matter that produces neutrinos, the answer depends on the
neutrino mean free path just outside the transition surface. (ii) How
much energy is deposited by neutrinos in the surrounding unconverted
matter? This depends on several factors such as the average neutrino
energy, the local temperature and density, and $Y_e$, the electron
fraction of the untransformed neutron star, all of which influence the
neutrino mean free path in normal matter. $Y_e$ and the temperature
depend in turn on the degree of deleptonization, or the time between
the formation of the neutron star and the transition to {\it (u,d)}
matter in the core. If the transition occurs in the protoneutron star
(PNS), $T\sim 50$~MeV and $Y_e\sim 0.3$. If the transition occurs well
after the birth of the NS, $T\sim 50$~keV, $Y_e\sim 0.01$, and
neutrinos free stream through the envelope.  (iii) How are the
neutrino rates affected if {\it (u,d,s)} matter is in a color
superconducting state?  This is relevant for the PNS since critical
temperatures can be as large as 50 MeV and neutrino emission and
absorption is known to be strongly modified in such phases.

Phase transitions into strange matter may cause mass ejection due to a
core bounce (Fryer \& Woosley, 1998; see also Takahara \& Sato, 1988
and Gentile et al. 1993). In this kind of model, a large
strange matter core (with a radius of 4-6~km) is formed and the
neutrinos are trapped long enough that they cannot efficiently
transport energy to the outer layers; in other words, the resulting
neutrino wind is not luminous enough to lead to mass ejection.
However, according to hydrodynamical simulations the core bounce may
cause baryon rich mass ejection with relativistic $\Gamma$-factors of
the order of $\Gamma\sim 40$, which is much lower than needed for a
gamma ray burst. In this work we study the other extreme and neglect
the core bounce. We show that in a hybrid star with a small initial
core (a radius of 1-2~km) of (u,d) quarks a phase transition into
strange matter can lead to a neutrino driven wind that can expel some of
the matter of the outer hadronic envelope of the star. A very small
core does not produce enough neutrinos to cause mass ejection, whereas
in case of a larger core neutrinos are trapped long enough that the
wind remains too weak to expel anything.  It would be interesting to
study both core bounce and the role of neutrinos simultaneously, but this is
left as an avenue for future work.

This paper is organized as follows: In Sect.~\ref{sec:quarknova} we
describe some important features of the QN. In
Sect.~\ref{sec:processes} we identify the dominant neutrino emission
and absorption processes and estimate relevant timescales for neutrino
diffusion. The full neutrino luminosity and the corresponding mass
ejection is calculated in Sect.~\ref{sec:lumin}.  We present our
conclusions and the scope of future investigations in
Sect.~\ref{sec:conclusion}.

\section{Quark Nova}
\label{sec:quarknova}

The basic idea in the picture we call QN is that the strange matter core, once
formed, will be isolated from the rest of the star. Charge neutrality
at the {\it (u,d,s)} surface requires the presence of electrons which
are bound to it by strong electrostatic fields that are determined
self-consistently in a Thomas-Fermi approximation (Glendenning et
al. 1995). This electron gas extends well outside the sharp surface of
the {\it (u,d,s)} object. The resulting potential difference is strong
enough to hold up protons against gravity, resulting in a
Coulomb gap of 200~fm (Alcock et al., 1986).  Neutral particles such
as the neutrons constituting most of the envelope/hadronic material
will traverse this gap, but protons and other positive ions will be
repelled by the enormous electric field. Following conversion to {\it
(u,d,s)} matter, a large flux of neutrinos is emitted which can
deposit energy in the surrounding matter, leading to a mass
outflow. Below, we present a first attempt to quantify this scenario.

\subsection{Dynamical timescales}

The phase transition {\it (u,d)} $\rightarrow$ {\it (u,d,s)} inside the 
HS propagates as a detonation roughly at the speed
of sound $c_s=c/\sqrt{3}$. 
A core few kilometers in radius transforms into a {\it (u,d,s)} object
within $\sim 0.1$ ms (Lugones et al., 1994). Although
weak processes can generate strange quark matter much faster than this, the
likelihood that it occurs simultaneously inside the whole object is vanishing.

Interestingly, the hydrodynamical
timescale $t_{\rm coll}$ for the shrinking of the core is within
the same order of magnitude as the free fall time scale,
 which is of the form (e.g. Chapter 18.5 in Shapiro \& Teukolsky 1983)
\begin{eqnarray}
t_{\rm coll} = \sqrt{\frac{1}{G\rho}}\simeq 0.1\, {\rm ms} \
\sqrt{\frac{10^{15}\, \rm g\, cm^{-3}}{\rho_{\rm uds}}}\ ,
\end{eqnarray}
which suggests that the conversion and shrinking of the
core occur almost simultaneously.

Given the jump in the density between the HS core and the hadronic
envelope ($\ge 2$), we expect the {\it (u,d,s)} core to shrink faster
than the envelope can respond. Roughly, $t_{\rm coll}^{uds}/t_{\rm
coll}^{env}\simeq \sqrt{\rho_{\rm env}/\rho_{\rm uds}}\simeq
\sqrt{0.5}$.  The spatial gap (see Figure~\ref{fig2}), given as
$\delta R/R_{\rm ud} = 1 -(\rho_{\rm ud}/\rho_{\rm uds})^{1/3}$, is
of the order of a hundred meters and is much larger than the Coulomb
gap, implying a large density discontinuity between hadronic and quark
matter. 

\subsection{Energetics}

Energy is released in both conversion of baryonic matter into 
strange matter in the form of latent heat and in the form of gravitational
potential energy.
The energy per baryon at zero pressure is speculated to be 50~MeV
lower in strange matter than in normal matter, and this energy can be
released in the phase transition to strange matter (see
e.g. Glendenning 1997, p. 351). Integrated over the entire neutron
star mass it would be roughly $10^{53}$~ergs. Moreover, the star
shrinks and the gravitational potential energy released is of the same
order.  As with core-collapse supernova, it is suggested that this
energy can be released in the form of neutrinos and/or in the form of
luminous ejecta (ODD) depending on the fate of the neutrinos once generated. 
Energy can also be released in the form of
gravitational waves (GWs), where the period of one pulsation is
$f=2\pi\sqrt{R^3/GM}\simeq 4\times10^{-4}$~s (Cheng \& Dai 1998).  It
is interesting to notice that the pulsation timescale defines the
collapse time scale (or $1/(4f)\simeq 1$~ms) if the conversion front
into strange matter propagates at least as fast as the star shrinks.
In this case, where the deconfinement transition occurs in a dynamical
timescale, the GW emission from the phase transition has
been examined by Marranghello, Vasconcellos, \& Pacheco (2002).
They found that even in the most favorable case corresponding
to rapidly rotating (ms period)\footnote{Recent studies have shown that young {\it (u,d,s)}
stars are stable to the viscosity-driven r-mode instability (Madsen 2000; Andersson, Jones, \& Kokkotas 2002).  These investigations
concluded that the instability cannot develop in {\it (u,d,s)} stars in
any astrophysically relevant temperature window (Gondek-Rosi\'nska, Gourgoulhon, \& Haensel 2003).} stars, only 1\% of the available energy will
be emitted as GWs.  These phase transitions excite mainly
the radial modes of the star (Sotani, Tominaga, \& Maeda, 2002)
which can only emit GWs when coupled with rotation (Chau, 1967).
As for non-radial modes, and extrapolating from studies done for  
 the case of newly born neutron stars (e.g., Ferrari, Miniutti, \& Pons, 2002),
these would have comparable frequencies
but damping timescales at least one order of magnitude higher than the
radial modes coupled to rotation; they would carry away
 a negligible amount of the energy released.  
To conclude, the bounds on the magnitude of GWs and on the 
amount of energy that they can take away during
a QN  suffer from uncertainties (e.g, values of the viscosity
of the {\it (u,d,s)} matter).
How these would exactly affect the dynamics of the QN event
is still uncertain at this stage. 
We now proceed to take a closer look at the neutrinos.

\section{Neutrino processes}
\label{sec:processes}

\subsection{Conversion in the core}

Once the {\it (u,d)} density is reached in the core of the HS (following spin-down or accretion), conversion into {\it (u,d,s)} matter
leads to neutrino emission via the following reactions (Iwamoto 1980; Dai et al. 1995):
\begin{eqnarray}
u+e^- & \leftrightarrow & d + \nu_e \, ,\nonumber \\
u+e^- & \leftrightarrow & s + \nu_e \, ,\nonumber \\
u+d   & \leftrightarrow & u + s \, ,\nonumber \\
d &\rightarrow & u + e^- +\bar{\nu}_e\, ,\nonumber \\
s & \rightarrow & u + e^- +\bar{\nu}_e \, .
\label{eq:reactions}
\end{eqnarray}

In case neutrinos are nondegenerate (low temperatures, when
neutrinos can escape), the rates for the first two reactions, which are identical to the rates of the fourth and fifth reactions respectively,
were derived by Iwamoto (1982) and Duncan et al. (1983). The first reaction
can proceed at an appreciable rate only if strong interactions are also taken into account, in which case it dominates the neutrino emissivity 

\begin{equation}
\epsilon_{q\beta}=1.9\times 10^{25}\left(\frac{n_b}{n_0}\right)T_9^6~{\rm erg~cm}^{-3}{\rm s}^{-1} \, ,
\end{equation}
where we have set the strong coupling constant $\alpha_s=0.1$ and
$Y_e=0.01$, typical of cold quark matter (Iwamoto (1982)), with
approximately equal number of $u,d,s$ quarks. $T_9$ is the temperature
in units of $10^9$~K and $n_B/n_0$ is the ratio of baryon density to
nuclear matter saturation density $n_0=0.16$~fm$^{-3}$. The rate for
the second reaction in eq.(\ref{eq:reactions}) is Cabbibo suppressed,
hence an order of magnitude slower, and the third reaction is
suppressed at tree level in the Standard Model, but can become
important at high temperatures (see below).

On the other hand, for temperatures of tens of MeV, at typical ${\it (u,d,s)}$ densities, the neutrinos are trapped and degenerate. Emission rates for
degenerate neutrinos were derived by Dai et al. (1995), 
which implies an emissivity
\begin{equation}
\epsilon_{q\beta}=2\times 10^{40}\left(\frac{n_b}{n_0}\right)^{2/3}T_{11}^3~{\rm erg~cm}^{-3}{\rm s}^{-1}\, , \label{epsdeg}
\end{equation}
where we have chosen $Y_e=0.3, Y_{\nu}=0.1$ (Prakash et
al (1995)), $\alpha_s=0.1$ and $T_{11}$ is the temperature in units of
$10^{11}$~K or $\sim 10$~MeV.  Note that the temperature dependence is
only $T^3$. In comparison, for the third reaction in
eq.(\ref{eq:reactions}), the numerical factor in front is 4 orders of
magnitude smaller. However, its rate is proportional to
$T^5$. Increasing the temperature by a factor of 2 increases the rate
by a factor of $2^5$, hence this process can become important only at
very high temperatures.

The emissivity estimated above does not reflect the true luminosity of the star
since neutrinos are trapped as we show in the following section. But first
let us estimate first order General Relativity effects on the neutrino emissivities. This can be introduced through redshift factors which
can be expressed in terms of the core radius $R_1$ since in
our case the core mass can simply be written as 
$M_{1} = 4\pi/3 \rho_{\rm uds}R_{1}^3$.
In Figure 3 we plot the ratio between the redshifted luminosity
(effectively seen by the envelope material) to the
 non-redshifted luminosity as a function of the core size
and for two different {\it (u,d,s)} density.
It is clear that a more compact core (higher
(u,d,s) density implying higher masses for a given
radius) will induce higher redshift.
In general cores smaller than 2 km induce almost no redshift.
Note that neutrinos, if not trapped, will be further redshifted
as they stream outwards since they are
subject to an increasing core size. Nevertheless we expect
at most the luminosity to be reduced by 50\% in our first estimate.  
While the gravitational redshift will degrade the total energy deposition above the core other effects related to bending of neutrino trajectories 
might compensate the loss (e.g., Cardall \& Fuller 1997).

\begin{figure}[t!]
\centerline{\includegraphics[width=0.5\textwidth]{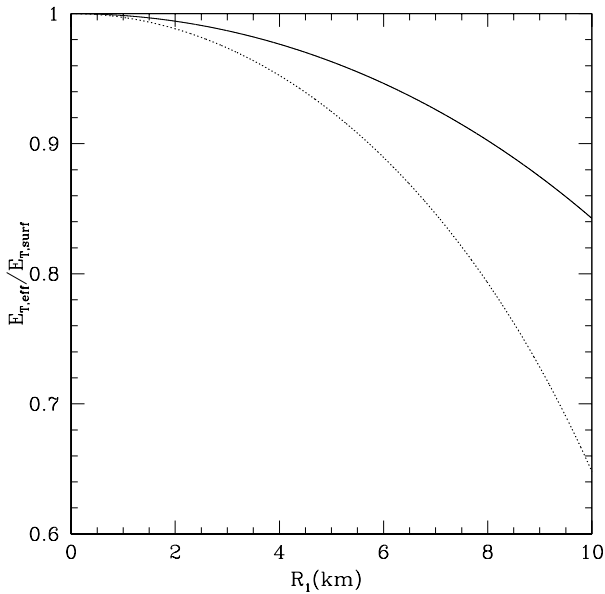}}
\caption{The ratio of the redshifted to the non-redshifted emissivity
versus core radius (reflecting the compactness parameter). 
The lower dotted curve
is for $\rho_{\rm uds}= 10^{15}$ g cm$^{-3}$
while the upper solid curve is for $\rho_{\rm uds}= 4\times 10^{14}$ 
g cm$^{-3}$.  In general, for cores smaller
than 2 km general relativistic effects are negligible.}
\end{figure}

\subsection{Diffusion timescales}

In this section, we present neutrino opacities that support the claim that neutrinos are essentially thermalized and degenerate in quark matter at MeV temperatures. Besides the inverse processes described in eq.(\ref{eq:reactions}), neutrinos can be absorbed by pair annihilation, 
\begin{equation}
\nu \bar{\nu} \rightarrow e^+ e^- \, .
\end{equation}
Even at $T\simeq 10$ MeV, where positrons can be created copiously, it is found that the cross-section for this process is 6 orders of magnitude smaller than neutrino scattering off electrons, hence this process can be ignored for estimating the opacity. Absorption by quarks dominates over neutrino-quark scattering at temperatures of tens of MeV and supra-nuclear densities (Steiner et al., 2001).

The main scattering processes are
\begin{eqnarray} 
e\nu& \rightarrow & e\nu \, , \nonumber \\
q \nu & \rightarrow & q \nu \, .
\end{eqnarray}
and the neutrino opacity is controlled by scattering against degenerate electrons, for which the cross section is given by (Burrows\&Thompson, 2002)
\begin{equation}
\langle \sigma \rangle  \simeq 
 2.5\times 10^{-42}\, {\rm cm}^2 \,\left( \frac{E_\nu(T_{\nu}+\mu_e/4)}{(10\, {\rm MeV})^2} \right)\ .
\end{equation}

The mean free path is then
\begin{equation}
\lambda  =  \frac{1}{n\langle\sigma\rangle} = \, 500\, {\rm cm} \, \left(\frac{(10\, {\rm MeV})^2}{E_{\nu}(T_{\nu}+\mu_e/4)}\right)\left(\frac{10^{15}\, {\rm g\, cm^{-3}}}{\rho}\right),
\end{equation}
The corresponding diffusion time we estimate is 
\begin{eqnarray}
\tau&=& (\Delta R)^2/(\lambda c)\simeq
0.1\, {\rm s}\\ \nonumber
 &\times&\left(\frac{\Delta R}{10\, {\rm km} } \right)^2
\left(\frac{\rho}{10^{15}\, {\rm g\, cm^{-3}}} \right)
\left(\frac{E_\nu(T_{\nu}+\mu_e/4)}{(10\, {\rm MeV})^2} \right)\, .
\label{eq:difftime}
\end{eqnarray}
Substituting typical numbers $E_{\nu}\approx T_{\nu}\simeq 10$ MeV, 
$\mu_e\approx 100$ MeV and $\rho=\rho_{u,d,s}\sim 10^{15}\, 
{\rm g\, cm^{-3}}$, neutrinos are certainly trapped long enough 
($\sim 10-100$ ms) to thermalize and to acquire a black body distribution 
defined by the {\it (u,d,s)} effective temperature.
Even the neutrinos produced in the outermost layers of 1~km (making up $\sim$~27\% of the total neutrinos) are diffused out in the timescale of the order of 1~ms. 
This implies that unless the size
of the {\it (u,d,s)} core is
small ($\Delta R << 10$ km), neutrinos are trapped
long enough to allow conversion of most of the 
 hadronic envelope material into {\it (u,d,s)}.

\subsection{Conversion of the envelope material}

 The infalling neutrons traversing the potential barrier will be
converted into strange quark matter via the reactions in
eq.(\ref{eq:reactions}). As mentioned previously, this can happen
before diffusing neutrinos reach the contamination front, unless the
core radius is small and neutrinos have an energy scale less than about 
 10 MeV. Note that since the neutron free fall time is much
smaller than its free decay lifetime, most neutrons reach the
contaminating front of ${\it (u,d,s)}$ before decaying to produce
neutrinos. Thus, neutrino emission in this time interval is
negligible. They will however be produced in flavor equilibrating
reactions in strange quark matter.  Since the timescale of weak
interactions is three orders of magnitude faster than the timescale of
neutron free-fall (or, the rate of shrinking of
the star), the weak interaction reactions have enough time to
equilibrate flavors as well as produce neutrinos (which can also
thermalize due to their short mean free path).

\section{Neutrino luminosity and mass ejection}
\label{sec:lumin}

For the case where most of the neutrinos are trapped (and
thermalized), in the first approximation, the neutrino luminosity can
be written as a black-body with the appropriate counting for fermion
statistics, leading to
\begin{eqnarray}
L_{\nu}= \frac{7}{8}N_{\nu} 4\pi R^{2} \sigma T_{\nu}^{4}&\simeq&
 3.2\times10^{53}\, {\rm erg\ s^{-1}}\\\nonumber
&\times& \left(\frac{R}{10~{\rm km}}\right)^2 \left(\frac{kT_{\rm eff}}{10~{\rm MeV}}\right)^{4}\, ,
\end{eqnarray}
where $T_{\rm eff}$ is the
core temperature and $N_{\nu}=3$ is the number of neutrino species.
This can be compared to the Eddington luminosity
\begin{eqnarray}
L_{{\rm Edd},\nu} &\approx& 3.0\times 10^{53} \,{\rm erg\ s^{-1}}\\\nonumber
&\times& \left(\frac{R}{10\ {\rm km}}\right)^3\left(\frac{\rho_{\rm uds}}{10^{15}\ {\rm g cm}^{-3}}\right) \left(\frac{10\, {\rm MeV}}{kT_{\nu}}\right)^{2}\, .
\end{eqnarray}
The above expressions show that $L_{\nu}> L_{\rm Edd}$ for $T_{\nu}> T_{\nu, c}=10$ MeV where the subscript {\it c} stands for critical.

The energy release in the phase transition is the same all over the core, 
so we can safely assume that the temperature 
is uniform and the average energy of neutrinos
in thermal equilibrium would be 
 $T_{\nu, equi.}\simeq 100/4$~MeV (since $n_u\simeq n_d=n_s\simeq n_\nu$).
Since $T_{\nu, equi.}>T_{\nu, c.}$, super-Eddington luminosities
are achieved. Nevertheless, according to eq.~(\ref{eq:difftime}), this
also implies much longer neutrino diffusion timescale and even
those generated in the outermost layers of the core are trapped
leading to the conversion of most of the envelope with little
or no neutron ejection by neutrinos.

\subsection{Ejection of neutrons}

In the event that a significant fraction of the neutrinos manage to
diffuse out of SQM before the conversion of the outermost layers of
the hadronic envelope (e.g., for $\rho_{uds}$ or
core size such that $\tau <
t_{coll}^{env}$ and small $E_{\nu}$), they will interact with the
electrons in the surface layer of quark matter and then with infalling
neutrons. The $\nu-$(free) neutron absorption cross-section is larger
by a factor 4 or more than elastic scattering at neutrino energies of
1 MeV, so one expects conversion of free neutrons to $p+e^-$. This
cross-section is quite large (about $10^{-39} {\rm cm}^2$, implying a
mean free path of a few hundred meters at envelope densities of
$10^{13}$g/cc).  These sources of opacity (neutrino scattering off
neutrons and electrons as well as absorption by nucleons) imply that
for 1~MeV neutrinos the mean free path is roughly $\lambda_{\rm env}
< 5\, {\rm km}$. Thus, it is possible that the neutrino energy is
dumped in lower density nuclear matter surrounding the quark core, which
we now proceed to estimate.

The total energy deposited by neutrinos in the nuclear envelope is given by
\begin{equation}
E_T=\tau\int_{R_1}^{R_2}Q_{\nu}(r)dr
\end{equation}
where $\tau$ is a typical time for free-fall collapse, given
approximately by $\tau=(R_2-R_1)/v_c$ and $v_c$ is the propagation
speed of the detonation wave (we may take it to be the speed of sound
in hot nuclear matter), $R_1$ and $R_2$ are the inner and outer radii
of the nuclear envelope, and $Q_{\nu}(r)$ is the energy deposition
rate per unit length at radius $r$. The energy loss is principally due
to inelastic scattering of neutrinos on nucleons, with a typical mean
free path of $\lambda_{\rm env}\simeq 1$ km.  As we are interested in an
order of magnitude estimate, we will approximate the radial dependence
of the emissivity as tantamount to absorption by a thick medium of
free scatterers, and indicate later how this approximation may be
improved. For the simple exponential profile of energy loss, the
emissivity from each successive volume element with increasing $r$
drops as $\epsilon_{\nu}(r)=\epsilon_{\nu}(R_1){\rm
e}^{-(r-R_1)/\lambda_{\rm env}}$.  It follows that
$Q_{\nu}(r)=4\pi~r^2\epsilon_{\nu}(R_1)(1-{\rm
e}^{-(r-R_1)/\lambda_{\rm env}})$. $\epsilon_{\nu}(R_1)$ is the
neutrino energy impinging per unit volume per unit time at $r=R_1$,
and is approximately equal to $L_{\nu}/(\frac{4}{3}\pi R_1^3)$,
neglecting for the moment energy loss in the thin crust above quark
matter. For the
fiducial values $R_1=2$ km ($\tau=1$ ms), $R_2=10$ km, $T_{\nu}\sim
T_{\rm eff}= 10$ MeV ($\lambda_{\rm env}=5$ m), we find
\begin{equation}
E_T=1.5\times 10^{51} {\rm ergs}
\end{equation}
Energy deposition effectively stops when the infalling nuclear envelope is as
thin as the typical mean free path. Our numbers can be improved with
a more detailed treatment of the radial dependence of the energy
deposition rate, and blast wave hydrodynamics. Although it appears that we have neglected neutrino emission processes from neutron matter itself, like the modified urca or neutrino bremsstrahlung, their inclusion does not change the order of magnitude estimate since for $T\leq 10$ MeV, the emissivity is comparable to or less than the emissivity from direct urca in quark matter. (This is because the $T^8$ dependence of these processes in neutron matter is more than compensated by smaller numerical pre-factors as compared to the direct urca in quark matter). We have also ignored
nucleon correlations in dense matter and Pauli blocking
effects, which effectively decrease the opacity, and would serve to soften the exponential fall-off of the emissivity. This implies that more energy could be deposited in the outer layers, resulting in somewhat increased mass ejection rates than we estimate below.   

Mass ejection occurs when energy deposited (heating) by neutrinos in a
shell of thickness $dr$ at radius $r$ exceeds the gravitational energy
density\footnote{ The material is heated primarily via the
charged-current absorption processes on free nucleon as $\nu_{e}
n\rightarrow p e^{-}$ and $\bar{\nu_{e}} p \rightarrow ne^{+}$.  A
particle is unbound/ablated if the sum of its energies is positive;
here we omitted the internal energy since previous studies have shown
that it has practically no effect on the total amount of ejected
material (Rosswog et al. 1999).  In our simplified model this implies
that most of the heat is converted to kinetic energy.  In contrast to
our estimate on the energy deposition by neutrinos, this introduces an
overestimation of mass ejection, so that these effects compete in
opposite directions.} induced
by the mass $M(r)=M_1+M_{\rm env}(r)$ with $M_{1}=4\pi/3~\rho_{\rm
uds}R_{1}^3$ as the core mass and $M_{\rm env}$ the envelope mass
contained within $r$,
\begin{equation}
f(r) = \frac{\epsilon_{\nu}(R_1)(1-{\rm
e}^{-(r-R_1)/\lambda_{\rm env}})~\tau}{\frac{G~M(r)}{r}\rho_{\rm env}(r)}
> 1\ .
\end{equation}

\begin{figure}[t!]
\centerline{\includegraphics[width=0.7\textwidth,angle=0]{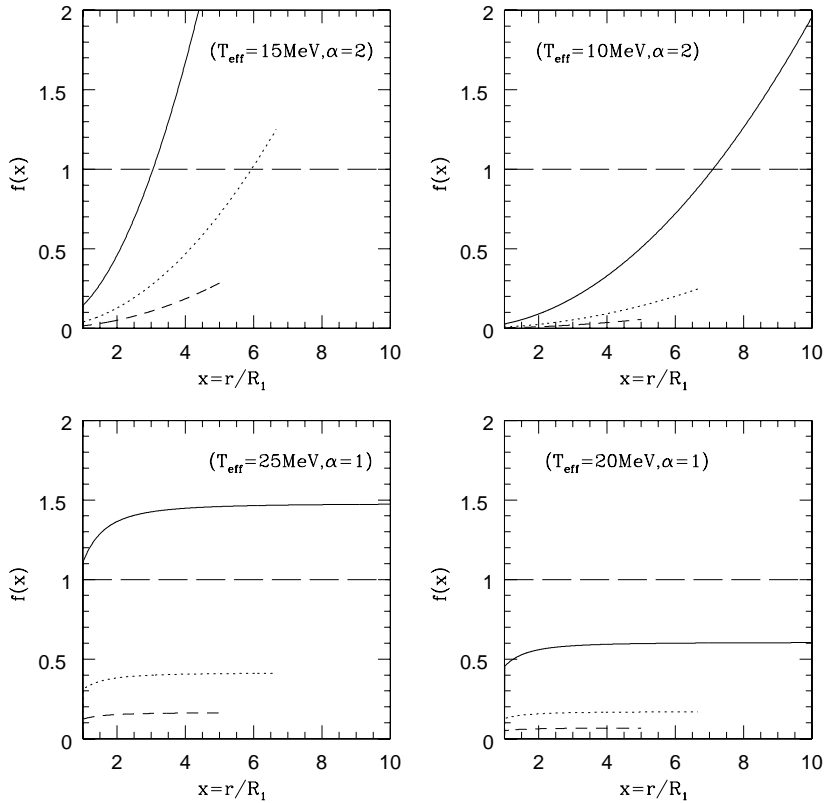}}
\caption{The ratio of neutrino energy density deposited to gravitational energy
density versus radius of the envelope ($x=r/R_{1}$) for
different core radius $R_1$. Mass ejection occurs
when $f > 1$. 
Two different envelope density distribution $\rho_{\rm env}\propto r^{-\alpha}$
are shown, 
$\alpha =2$ (upper panels) and $\alpha=1$ (lower panels).
}
\label{fig3}
\end{figure}

\begin{figure}[t!]
\centerline{\includegraphics[width=0.5\textwidth,angle=0]{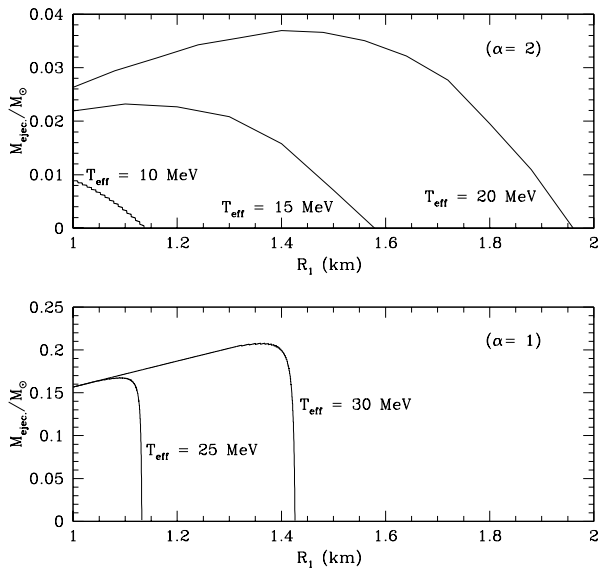}}
\caption{The total mass ejected (in solar units) versus the core size ($R_{1}$) for
different core temperatures ($T_{\rm eff}$). 
Two different envelope density distribution $\rho_{\rm env}\propto r^{-\alpha}$
are shown, 
$\alpha =2$ (upper panel) and $\alpha=1$ (lower panel).
}
\label{fig4}
\end{figure}

Figure~\ref{fig3} shows the ratio ($f(x)$) between these two
components versus radius ($x=r/R_{1}$) for different core sizes
($R_{1}$).  
Figure~\ref{fig4} shows the total mass ejected versus
core radius $R_{1}$ for different core temperature, $T_{\rm eff}$.
While ejection of inner layers (satisfying the $f(r) > 1$ condition)
might be prevented by the overlaying envelope matter, for the cases we studied
we find $f(r>r_{\rm min})>1$; $r_{\rm min}$ corresponds to 
 $f(r=r_{\rm min})=1$. Thus the total ejected
mass can simply be expressed as $M_{\rm ejec}=\int^{R_2}_{r_{\rm min}}
4\pi r^2 \rho_{\rm env}(r) dr$.
We used our fiducial values, $R_{2}=10$ km, 
 $T_{\nu} = T_{\rm eff}$ and $\rho_{\rm env}/\rho_{\rm uds}
=1/2$. We parametrize the density variation in the neutron rich
envelope as $\rho_{\rm env.}\propto r^{-\alpha}$; we consider the
$\alpha=1$ and $\alpha=2$ cases which reasonably describes the density
variation over a radius range from 2-10 km, gathered from studies of
neutron skin thickness in Pb nuclei (Horowitz \& Piekarewicz 2001), as
well as prior studies of the neutron matter equation of state (Strobel
et al. 1997).

As can be seen in 
 Figure~\ref{fig3} and Figure~\ref{fig4} the $\alpha =1$ case requires extreme
conditions (very small cores and high effective temperatures)
for heating to ablate material. These extreme
conditions also imply extreme mass ejection since the $\rho_{\rm env}
\propto r^{-1}$ profile provides enough material in the outer layers.
Specifically, the $\alpha=1$ case requires temperatures
above the 20 MeV range and core radius $R_1 < 1.2$ km
for ejection to occur. Whether such small cores
can provide the extreme temperatures following the
collapse is questionable.
As for the $\alpha=2$ envelopes, mass ejection is triggered
for core temperatures as low as 10 MeV. For example,
for a core temperature $T_{\rm eff}\le 15$ MeV, a
1.5 km core can ablate up to $0.01M_{\odot}$ of envelope material.  

We note that for the same envelope
mass, $\alpha=1$ envelopes are less dense than
the $\alpha=2$ ones. The neutrinos deposit less
energy in such envelopes (larger $\lambda_{\rm env}$)
explaining the need for extreme temperatures for ablation
to be triggered. In both cases, however, 
we find that for small cores the neutrino flux 
is too small to account for any mass ejection while larger cores 
 trap neutrinos long enough for the entire envelope to be converted
to {\it (u,d,s)} matter. 
Finally, since the $\rho_{\rm env} \propto
r^{-2}$ dependency is reflective of the outer radii 
of the neutron-rich envelope this favors the scenario where 
only the outermost layers will be ejected in QNe.
Whether this ejected mass succeeds in turning
into a wind and escape to infinity, or
is later prevented by the fallback
material remains to be confirmed.

\subsection{Ejection of crust material ?}

The crust material ($\rho_{\rm crust} < 10^{11}\, {\rm g\, cm^{-3}}$)
subject to the Coulomb gap would remain suspended above the {\it
(u,d,s)} and would most certainly be subject to the delayed neutrino
burst. The amount of the crust ejected depends on the energy
deposition rate which is determined by the crust opacities; neutrino
scattering off nucleons, nuclei and electrons as well as absorption by
nucleons and nuclei all contribute, and the total opacity in the crust
averaged for all neutrinos and anti-neutrinos is (Lamb \& Pethick 1976)
\begin{equation}
\kappa_{\rm t}\simeq 1.2\times 10^{-6} \left(\frac{\rho}{10^{10}\, {\rm 
g\ cm^{-3}}}\right) \left(\frac{kT_{\nu}}{10\, {\rm MeV}}\right)^2\, {\rm 
cm^{-1}},
\end{equation}
implying a mean free path $\lambda=1/\kappa_{\rm t}$ that is several
orders of magnitude larger than the crust thickness.  It is thus very
unlikely that the crust will be blown off by neutrinos unless photons
are at play as might be the case if color superconductivity sets in
the core, (Ouyed \& Sannino, (2002)). This also justifies neglecting the 
crust for the estimate of energy deposition in nuclear matter in the 
previous section.

\subsection{Effects of color superconductivity}
 
Due to the large critical temperatures associated to color
superconductivity, the core may enter the color-flavor-locked (CFL)
phase, with the surface in the 2-flavor superconducting phase
(2SC). In that case, neutrino emission is dominated by the decay and
scattering of collective excitations, rather than free quarks. The
neutrino opacity in the temperature range 10-30 MeV is a few meters,
determined by neutrino absorption on the massless mode that describes
baryon superfluidity (Reddy et al (2003)). For $T\sim 10$ MeV (still
much smaller than the gap), the emission rates in CFL-paired matter
are roughly three orders of magnitude smaller than unpaired quark
matter, although the opacities are similar due to efficient scattering
off Goldstone modes.  Therefore, the neutrino flux from CFL matter is
only about 1\% of unpaired quark matter. Since the envelope is
composed only of normal matter, a similar suppression is expected in
the total mass ejected.  Although the emission rate from the 2SC phase
is only slightly reduced compared to normal quark matter, (beta decay
of free light quarks of one color dominates), the fact that it occurs
in a surface layer implies that the integrated luminosity receives
contributions predominantly from CFL matter.  The absence of Goldstone
modes in the 2SC phase implies reduced opacities, so that the
neutrinos from CFL quark matter can escape once they reach the quark
matter surface. For temperatures of the order of the superconducting
gap $\Delta$, quasi-free excitations (Pair-breaking) can become
dominant sources of neutrino emission and scattering (Jaikumar,
Prakash, \& Schaefer 2002; Kundu \& Reddy 2004). The dominant
neutrino rates then approach those in normal matter. Therefore, we
conclude that the neutrino flux from the quark core, and hence the
mass ejection is suppressed by at most 99\% for $T\ll\Delta$
(effective quenching), and exponentially approaches the rate computed
with normal quark matter as $T\rightarrow T_c$ (since the
suppression of direct urca goes like exp(-$\Delta/T$)).  In light of
the uncertainty in the exact densities at which different
superconducting phases appear, and the value of the gaps themselves,
these estimates are reasonable.  The general conclusion is that the
transition to a superconducting state will diminish the amount of
energy deposition and mass ejection in the envelope through neutrinos.

\subsection{Neutrino oscillations and matter effects}

Recent experimental neutrino data provides evidence for
neutrino oscillations (for a recent review, see e.g. Maltoni et al.,
2004). Inside the dense core of the hybrid star, electron
neutrinos (and antineutrinos) created in processes described by eqs.~(2) 
are created as weak interaction eigenstates, but they propagate as 
the eigenstates of Hamiltonian, providing the mechanism for neutrino
oscillations. Matter affects neutrino oscillations (Wolfenstein 1978, 
Mikheyev \& Smirnov 1985) and may induce conversions between the neutrino 
states. These conversions occur at the so-called
resonance densities (see e.g. Dighe \& Smirnov, 2000). In the standard
picture, there are two such resonance regions with densities of roughly 
$1000-10000$~g/cm$^3$  and $10-30$~g/cm$^3$. These densities are much lower
than in the hadronic envelope, and therefore neutrino oscillations in
matter do not influence our estimates of mass ejection. Electron
neutrinos (and anti-neutrinos) propagate within the quark star
effectively in the same state that they were created.
   
In non-standard neutrino theories one may expect different effects. E.g. 
sterile neutrinos could, if created in conversion processes, carry away a
significant amount of explosion energy and influence the dynamics of
the QN. This kind of neutrino models have been extensively studied in
connection with supernovae, and it is found that sterile neutrinos with
masses in the keV-range may lead to conversions, but very
heavy or light sterile neutrinos have no resonances inside the core or
envelope (see e.g. Abazajian, Fuller \& Patel (2001) and Ker\"anen, 
Maalampi, Myyryl\"ainen \& Riittinen (2004)). We do not consider the 
possibility of sterile neutrinos further in this work.

To conclude, it appears that neutrino oscillations, masses and so-called 
matter effects do not change our results in the new standard neutrino 
physics picture. 

\section{Conclusion}
\label{sec:conclusion}

We have studied the role of neutrinos in a quark nova, and found that
they could deposit sufficient energy in the outer layers of the star
to cause significant mass ejection of the nuclear envelope. This
conclusion assumes a scenario in which the size of the
quark core ($R_{\rm core} \le 2$ km), is such that neutrinos can
diffuse out and lose their energy in the nuclear envelope well before
$(u,d,s)$ conversion followed by shrinking of the star is
completed. The mass ejection fraction of the outermost layers of the
envelope is estimated using the energy deposition rate by
neutrino-nucleon inelastic scattering. We predict on average up to
$\sim 10^{-2} M_{\odot}$ of neutron rich material to be ejected during the
explosion. This is suggestive of some interesting astrophysical
 implications such as r-process products injected into the
inter-stellar medium (e.g., Freiburghaus et al. 1999
and references therein) and neutron-rich disk forming around newly born
quark stars (e.g. Ker\"anen \& Ouyed 2003). 

On the other hand, for the case when ($R_{1} >> 1$ km) the conversion of 
the core into strange matter and the shrinking timescale into a dense 
quark object are faster than the neutrino diffusion time scale. The entire 
neutron star converts into a {\it (u,d,s)} object with no mass ejection.  

Although we have mentioned two distinct physical scenarios, one where
the envelope-core boundary is continuous, and another where they are
separated by a macroscopic distance, the estimates and conclusions on
neutrino transport and mass ejection are unaffected by this
difference. However, they must be distinguished since the latter involves
the interesting possibility of a low density region developing inside the star!
 
Clearly, to understand the complex energetics and dynamics involved in
the QN explosion and the consequences on the surrounding environment,
one needs the help of advanced numerical simulations where general
relativistic effects can also be taken into account. Simplifying
assumptions such as the stasis of the envelope while ejecting mass
can be relaxed. The radial dependence of the energy deposition by
neutrinos can be computed more accurately using Boltzmann equations
for neutrino transport. In this work, we have outlined the basic
physical picture of the quark nova and shown that mass ejection is
feasible, based on dominant neutrino emissivities and opacities. Our
future investigations are directed towards making this statement
quantitatively precise by including the refinements of blastwave
hydrodynamics and neutrino transport.

\begin{acknowledgements}

We thank E. Olsson and C. J. Pethick for helpful discussions and
suggestions. 
P.K. is grateful for the University of Uppsala for
the cordial atmosphere during the early stages of this work.
The research of R.O.  is supported by grants from the
Natural Science and Engineering Research Council of Canada
(NSERC). P.J. is supported in part by NSERC and in part by the Fonds
Nature et Technologies de Qu\'ebec.

\end{acknowledgements}

\end{document}